# High Precision Arithmetic for Scientific Applications

Foster Morrison

Turtle Hollow Associates, Inc.

Gaithersburg, MD, USA

Abstract

All but a few digital computers used for scientific computations have supported floating-point and digital arithmetic of rather limited numerical precision. The underlying assumptions were that the systems being studied were basically deterministic and of limited complexity. The ideal scientific paradigm was the orbits of the major planets, which could be observed with high precision, predicted for thousands of years into the future, and extrapolated for thousands of years into the past. Much the same technology that has made computers possible has also provided instrumentation that has vastly expanded the scope and precision of scientific analysis. Complex nonlinear systems exhibiting so-called "chaotic dynamics" are now fair game for scientists and engineers in every discipline. Today it seems that computers need to enhance the precision of their numerical computations to support the needs of science. However, there is no need to wait for the necessary updates in both hardware and software; it is easy enough to monitor numerical precision with a few minor modifications to existing software.

## 1. The State of Scientific Computing

Most, if not all, general purpose scientific computers support binary digital integer and floating-point arithmetic. The set of integers is infinite, whether one includes all of them or only the non-negative ones. It also is discrete; there are no integers less than 3 and greater than 2. Floating-point numbers are intended to serve the role of real numbers in scientific calculations, but they are, by necessity, a finite subset of the rational numbers. What all this means is that no calculation ever follows a mathematical model *exactly* except in some extremely trivial examples. Therefore any endeavor in engineering or science should include error estimates for the calculations and likely will necessitate efforts at error control.

Error control and estimation currently is strongly constrained by the limited range of numbers supported by the built-in arithmetic processing units of even the most powerful supercomputers. Floating-point numbers are supported by most scientific programming software to a precision of only 64 bits. The X86 CPUs in current PCs (personal computers) actually support 80-bit floating-point arithmetic, no more, no less, but most software truncates that to 64 bits. I am aware of this because my Lahey Fortran[1] compiler offers the option of not truncating the 80-bit numbers as long as they are in the CPU, but the number is truncated when it is written to the computer memory. I believe the Motorola processors formerly used in Apple computers supported 96-bit floating point, but I am not aware of any software that took advantage of that capability.

Those of us who have worked with computers since the days of mainframes that filled large, powerfully air-conditioned rooms, recall that 64-bit floating point was the usual standard, since the word length on IBM computers was 32 bits and that was rarely enough even in those ancient days. The IBM 709 was an advanced (for its day) computer powered by vacuum tubes, a thing which younger scientists and engineers probably never have seen. The design was updated with transistors to the IBM 7090 and 7094, which proved very popular. Some manufacturers, such as Honeywell and CDC implemented longer word lengths, which allowed many programs to use an extended single precision longer than 32 bits, but shorter than 64, for most applications. With the introduction of the 360 series computers, IBM introduced memory consisting of 8-bit words, called "bytes," and the floating-point arithmetic became standardized to 64-bit "double precision" among all software producers, though some CPU producers did offer slightly more powerful options.

This byte-oriented architecture carried over to a new market in minicomputers, a market dominated by the Digital Equipment Corporation from the 1960s to the 1990s.[2] The development of integrated circuits made the PC possible, so minicomputers lost their niche and disappeared. Of course, this technology was adapted to a new generation of mainframes, most notably supercomputers that now can execute quadrillions of floating-point instructions per second.[3] What is strange is that these high powered machines support the same modest level of floating-point precision as the ancient mainframes of the 1960s, namely 64 bits. In some cases this may not make any difference, since the computation might be for an immensely complex statistical simulation that samples a lot of things that are vaguely defined. But whatever the problem at hand, there should be some control over error estimation and error growth, which, it seems likely, is some kind of "random walk."

There are many software packages that do support extreme levels of numerical precision or even *accuracy*, sometimes to the point of being exact where that is possible.[4] However, this software is slow, since it is not supported by special hardware designed for rapid numerical computations. Obviously what is needed is hardware that can rapidly perform floating-point computations of extended precision levels well beyond 64 bits. And compilers that make it easy to use.

Of course, we will never be able to do the arithmetic of real numbers except for the special cases where mathematical analysis can supply the results:

$$\sqrt{2} \times \sqrt{2} = 2 \tag{1}$$

$$\sin(\pi/4) = \sqrt{2}/2 \tag{2}$$

There is a lot of mathematical analysis far more complex than this high school stuff, but not nearly enough to do what scientists and engineers want. It's pretty much a case of "You have to go to the computer because even elliptic functions can't do the job."

Now is there a simple example for the need for flexible, high precision floating-point computations? One possibility is to use the high precision, specialized software that mathematicians love so dearly. A quicker and easier approach is to use a truncation function that demonstrates what using less that 64-bit floating-point arithmetic can do. Then a simple demonstration can be given with an Excel spreadsheet using ROUND($x$, $n$), where $x$ is a number and $n$ a number of decimal places. Actually the arguments of the function are spreadsheet cells containing the numbers desired. Negative values of $n$ allow one to round to the nearest value of $10^{-n}$, e.g., ROUND(1234, -2) = 1200.

A well-known test case is the Lorenz equations that, for a proper choice of parameters, exhibit chaotic behavior.[5] These are three nonlinear ordinary differential equations that launched a new era of scientific and mathematical awareness often called "Chaos Theory." The humanities are filled with examples of chaos and endless studies of them, but scientists did not like anything for which they could not exhibit some mathematics. Prof. Lorenz, a meteorologist at MIT, filled this void while trying to construct a very simple model of atmospheric phenomena using an early model digital computer.

## 2. Demonstration Computations

The extended precision software mentioned above – with more than 16 significant decimal digits (or an approximate binary equivalent) – is currently available to a limited extent in something designed mostly for the efforts of theoretical mathematicians to do things like search for ever larger prime numbers, which theory asserts, exist in unlimited quantities.[6] However an adequate demonstration of the limitations of 64-bit "double precision" can be done rather easily with Excel spreadsheets.

The trick is to select a simple mathematical computation that is deterministic in one sense, but not in every sense. One such is a rotation, which can be expressed as a simple vector-matrix formula as

$$\mathbf{v}_1 = \mathbf{R}\mathbf{v}_0 \tag{3}$$

The symbols $\mathbf{v}_1$ and $\mathbf{v}_0$ represent column vectors of equal dimension $m$ and $\mathbf{R}$ is an $m$×$m$ matrix with the properties $\det(\mathbf{R}) = +1$ and $\mathbf{R}^{-1} = \mathbf{R}^T$. Rotation matrices of any dimension are possible, but those of dimensions 2 and 3 are most common, since they are fairly easy to visualize and adequate for most applications. For our test computations we will make use only of the 2-dimensional case, since it is easy to visualize and supplies all the dynamics needed.

What dynamics? Rotations can define a linear difference equation simply by generalizing (3) to

$$\mathbf{v}_{k+1} = \mathbf{R}\mathbf{v}_k, \ k = 0, 1, 2, 3 \ldots \tag{4}$$

The geometry is simple enough: going around in circles. It also is deterministic, in that every step is simply an integer multiple of the first one. In other words, if the initial rotation is 5°, the second merely increases that to 10°; then the sequence continues as 15°, 20°, 25°, &c. However, if the rotations are computed numerically, the vectors will drift from the theoretical values, since the linear difference equation (4) is not stable, despite being deterministic. All linear difference and differential equations are unstable, so it should surprise no one that "life can be so nonlinear." Such systems either collapse to a point or shoot to infinity or do one in some dimensions and collapse in others. However, they have numerous applications, such as teaching difference (and differential) equations to beginners (mostly undergraduates) and providing the means to do adjustments to both initial conditions and parameters in the nonlinear systems that are needed for so many scientific applications.

In the case of 5° rotations, one can solve (4) by multiplying to get a rotation of $5k$ degrees and then compute sines and cosines of that angle. Or one can multiply **R** numerically $k$ times. Since the sines and cosines of most angles of integer numbers of degrees are irrational, the computation of $\mathbf{R}^k$ will drift away from its theoretical value, albeit slowly if lots of significant digits are used. Eventually the vectors $\mathbf{v}_{k+1}$ will collapse or blow up. These simple computations provide a means to test the effects of various levels of numerical precision in this simple, seemingly deterministic geometric problem.

The lesson is that no level of numerical precision is always enough. But for the complex, nonlinear dynamical systems that describe what we think is "reality", there are no such precise, if not exact, benchmarks. Multiple numerical integrations must be run at various levels of numerical precision and perhaps with different numerical algorithms as well. Currently this is not feasible, even in the cases where 64-bit floating-point arithmetic suffices. The only readily available alternative is 32-bit arithmetic and that only rarely will prove to suffice.

Running a computation at a slightly reduced level of precision can be accomplished by inserting a rounding function at critical points. This would seem to be a good idea, especially since the speed of computers has advanced markedly during the past 50 years, but 64-bit floating-point arithmetic (about equivalent to 16 significant decimal digits) has remained the industry standard and what is supported by most software. Actually, the X86/87 CPU architecture almost universal in personal computers and similar devices supports 80-bit floating-point, but most software has truncated it to 64-bits. Some current Fortran offerings from Lahey support the 80-bit format and a few even add a 128-bit option supported by bit-by-bit manipulations.[7]

## 3. Sample Numerical Tests

It would be easy enough to code the rotations in Fortran or some other complier, like C, but is even easier to do in Excel spreadsheets. These even include the ROUND(…) function described above. This was done and several cases were examined. What we present here will be for 5° rotations, which go through a full circle in 72 steps. This is very neat, since the exact results can be computed for any number of steps that are an integer multiple of 18 or 36, as well as 72. Any others can be computed to the accuracy provided by subroutines for sines and cosines. For integer multiples of 90°, of course, these are -1, 0, or 1, so all the rotations are trivial and can be computed *exactly*.

The initial condition was $\mathbf{v}_0 = (1, 0)^T$. The length $\|\mathbf{v}_k\|$ was tracked, since the expected behavior is for it to collapse to 0 or soar to infinity eventually. How fast it does this is a measure of the effect of accumulated numerical errors. This can be done with truly useful mathematical models, such as nonlinear differential equations, in exactly the same way. An easy way to do this would be to insert the ROUND(…) function into the numerical integration algorithm at the end of each step and run the computation at least twice and preferably 3 or more times. This can provide some quality control even with the current widely supported standard of 64-bit floating-point numbers.

Comparisons were made of rotations computed with 16-digit numbers and 12-digit and 7-digit ones. After 4 full rotations (288 matrix multiplications) the length of $\mathbf{v}_{288}$ had risen by about 5.77E-15 for the 16-digit computation, 8.16E-11 for the 12 digit case, and dropped -1.07E-6 for the 7-digit ones. So it seemed that the higher precision calculation would explode after a very slow start, while the reduced precision one would gradually collapse. That it could go either way is to be expected.

Evidently it would take a good many rotations to lead to numerical destabilization, so the calculation was accelerated by repeatedly squaring the multiplication matrix thusly

$$\mathbf{R}_0 = \mathbf{R}$$

$$\mathbf{v}_1 = \mathbf{R}_0 \mathbf{v}_0 \qquad (5)$$

$$\mathbf{R}_{k+1} = \mathbf{R}_k \mathbf{R}_k, k = 0, 1, 2, 3 \ldots$$

$$\mathbf{v}_{k+1} = \mathbf{R}_k \mathbf{v}_k$$

This skips a lot of intermediate steps in the quest to reach obvious instability and, it should be noted, the results are identical in terms of analytic algebra. But algebra is one thing and floating-point arithmetic is another. So one can be reasonably certain that (5) will not give the same exact numerical results as (4), even for fairly small values of *k*. However, the goal is to demonstrate that (4) and (5) will become unstable more rapidly with diminishing numerical precision. This is fairly obvious, but the goal is to provide a simple numerical example that will illustrate how the instability can arise rapidly.

Figure 1 shows how the length of $\mathbf{v}_k$ in (5) grows from 1 with *k*. Those equations, of course, were designed to exhibit the numerical instability rapidly. The computations in (4) would get there at a more leisurely pace, but get there they would.  Note that the length began to drop using (4) with 7 significant decimal digits.  It continued to decline to 0 with ever more multiplying by the same matrix.  Evidently this property was not shared by the powers of numerical **R** used in (5).

In serious scientific computations the mathematical model will almost always be nonlinear, maybe a little bit, or maybe so much as to earn the designation "chaotic."  In such cases there is no nice analytic model to use to check the computations.  The days of perturbation theories and low precision data are long gone.  What lingers from the past is numerical calculations of significantly uncertain precision.

## 4. Conclusions

The 80 and 128-bit formats supported by select Lahey Fortran products, or whatever other such capabilities available, should be utilized for massive computations and maybe some that are not so massive.  The orbits of the major planets, and even those of artificial satellites, do not exhibit chaotic behavior, but those of asteroids may.  As long as they stay between Mars and Jupiter, there are no serious hazards.  But there are a few that come close to Earth and some astronomers and NASA have made serious efforts to track them.  It seems likely that a strike by a good sized asteroid caused the extinction of the dinosaurs.  That perhaps allowed humanity to come into existence, but the same sort of event could trigger human extinction or other inconveniences.

The previously mentioned Lorenz equations are highly simplified models of an atmospheric convection cell.  Chaotic behavior is quite common in meteorology and many other areas of scientific inquiry.  The obvious conclusions are that extended precision floating-point should have been implemented a long time ago, along with tests for numerical precision.

Considerable time will be required before something like open-ended floating-point numerical precision is made available for the rapid, massive computations needed for contemporary science.  As noted above, however, scientists and engineers can estimate numerical precision with minor modifications to existing software.  This, in fact, is how Prof. Lorenz discovered the chaos that a few people like Poincaré knew was there.[8]

It is not yet clear what level of precision should be supported by computer hardware (math co-processors).  Perhaps there is a limit to the number of variables and numerical precision that can be done with mathematical modeling, whether it be practical or theoretical.  But it is safe to say that no one yet knows what these limits are or might be.  It is long past the time to find out what is what.

## 5. Notes & References

1. *F77L Reference Manual*, Lahey Computer Systems, Inc., Incline Village, NV, 1988.

2. Digital Equipment Corporation, Wikipedia,
    http://en.wikipedia.org/wiki/Digital_Equipment_Corporation

3. Supercomputers, Wikipedia.
    http://en.wikipedia.org/wiki/Supercomputer

4. Arbitrary-precision Arithmetic, Wikipedia.
    http://en.wikipedia.org/wiki/Arbitrary-precision_arithmetic

5. Morrison, F., *The Art of Modeling Dynamic Systems*, Dover Publications, Mineola, NY, 2008, pp. 261-267, 344.

6. Rademacher, H. and O. Toeplitz, *The Enjoyment of Mathematics*, translated by H. Zuckerman, Princeton University Press, Princeton, NJ, 1957, pp. 9-13, 135-139.

7. E-mail from Lahey, Aug. 22, 2013; see:
    http://www.polyhedron.com/pb05-win32-language0html
     and http://www.polyhedron.com/pb05-linux-language0html

8. I recall this from a lecture that Prof. Lorenz gave many years ago at what is now NIST.

## Contact information

Foster Morrison
Turtle Hollow Associates, Inc.
PO Box 3639
Gaithersburg, MD 20885-3639
USA
Turtle_Hollow@sigmaxi.net

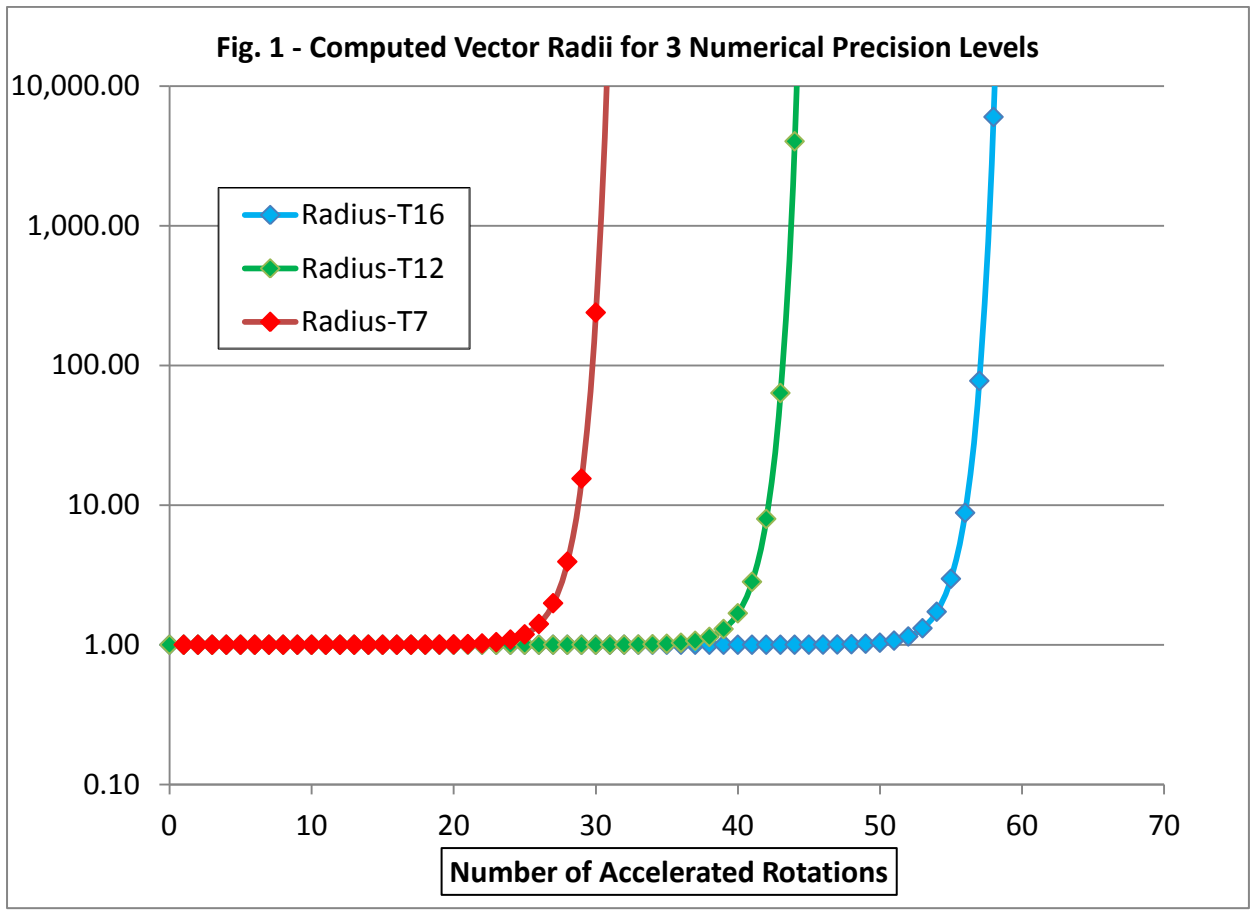